\documentclass[final,3p,times,twocolumn]{elsarticle}
\usepackage{graphicx}
\usepackage{amsmath}
\usepackage{hyperref}
\usepackage[version=3]{mhchem}
\usepackage[usenames]{color}

\renewcommand{\dj}{d\kern-0.4em\char"16\kern-0.1em}
\renewcommand{\DJ}{\raise0.3ex\hbox{-}\kern-0.36em D}

\usepackage{lineno}

\journal{Journal...}

\begin{document}
\begin{frontmatter}
\title{Study of the local field distribution on a single-molecule magnet--by a single paramagnetic crystal; a DPPH crystal on
the surface of an Mn$_{12}$-acetate crystal}

\author[Rudjer]{Dijana \v{Z}ili\'{c}}

\author[Rudjer]{Boris Rakvin}

\author[FSU]{Naresh S. Dalal}

\address[Rudjer]{Department of Physical Chemistry, Ru\dj er Bo\v{s}kovi\'{c} Institute,
\\Bijeni\v{c}ka cesta 54, 10000 Zagreb, Croatia}

\address[FSU]{Department of Chemistry and Biochemistry, Florida State University,
\\Tallahassee, FL 32306, USA}
\begin{abstract}
The local magnetic field distribution on the subsurface of a
single-molecule magnet crystal, SMM, above blocking temperature
($T >> T_b$) detected for a very short time interval
($\sim10^{-10}$~s), has been investigated. Electron Paramagnetic
Resonance (EPR) spectroscopy using a local paramagnetic probe was
employed as a simple alternative detection method. An SMM crystal
of
[Mn$_{12}$O$_{12}$(CH$_3$COO)$_{16}$(H$_2$O)$_4]\cdot$2CH$_3$COOH$\cdot$4H$_2$O
(Mn$_{12}$-acetate) and a crystal of 2,2-diphenyl-1-picrylhydrazyl
(DPPH) as the paramagnetic probe were chosen for this study. The
EPR spectra of DPPH deposited on Mn$_{12}$-acetate show additional
broadening and shifting in the magnetic field in comparison to the
spectra of the DPPH in the absence of the SMM crystal. The
additional broadening of the DPPH linewidth was considered in
terms of the two dominant electron spin interactions (dipolar and
exchange) and the local magnetic field distribution on the crystal
surface. The temperature dependence of the linewidth of the
Gaussian distribution of local fields at the SMM surface was
extrapolated for the low temperature interval (70--5~K).

\end{abstract}

\end{frontmatter}


\section{Introduction}

Single-molecule magnets (SMMs) are candidates for many
applications, such as quantum computation, high-density magnetic
data storage and magnetoelectronics
\cite{Leuenberger2001,Leuenberger2003}. In order to develop these
applications, it is important to investigate what happens when
they are in contact with other dissimilar materials, as it is
expected that their wave functions and/or magnetic fields extend
considerably outside the physical structure
\cite{Eisenmenger2003}. This also implies improved detection of
the magnetic fields on their surfaces \emph{i.\,e.\,}, the
magnetic field generated by a crystal in the vicinity of its
surface. The magnetic properties of SMMs at temperatures above
blocking temperature, $T>T_B$, can be considered as
superparamagnetic properties. The relaxation of magnetization in
this temperature region can be described by the Arrhenius law
$M(t)=M_0\exp(-t/\tau)$, with a characteristic relaxation time,
$\tau$ \cite{Gatteschi2006}. Such a relaxation time in the simple
form is given as a function of the energy barrier, $E_b$, and
temperature ($\tau =\tau_0 \exp(E_b/kT)$). When a magnetic field
is applied, the total energy barrier is expected to be lower
\cite{Gatteschi2006}. Thus, at the surface of the SMM crystal one
expects time-dependent as well as magnetic-field-dependent
distribution of the magnetic fields. In superparamagnetic systems,
the observed magnetic behavior strongly depends on the value of
the measuring time, $t_m$, of the employed experimental technique
with respect to intrinsic $\tau$. Therefore, wide variation can be
expected in the employed time ``window''-which varies from large
values as in magnetization measurements (typically 100~s) to very
small ones, as in ac susceptibility \cite{Blinc2003} or
spectroscopy \cite{Rakvin2005} ($10^{-8}$~s). The detection and
description of such distributions of local fields at the surface
as functions of the orientation of the SMM crystal in the external
magnetic field within a very short time interval ($\sim
10^{-10}$~s) are the focus of the present study.

Electron Paramagnetic Resonance (EPR) spectroscopy with a local
paramagnetic probe will be employed as an alternative simple
detection method. As shown recently
\cite{Rakvin2005,Rakvin2004,Rakvin2003}, an EPR technique in
combination with several characteristic paramagnetic probes shows
a reliable capability for such measurements for the most common
EPR spectrometers involving magnetic field $\sim$0.35 T with
corresponding X-band frequency ($\sim$9.5~GHz). An additional
characteristic of this detection is related to the possibility of
deducing a time-dependent fluctuation of the local field in a
short time scale \cite{Rakvin2005}. In the previous studies, the
probes used were a powder of 2,2-diphenyl-1-picrylhydrazyl (DPPH)
and needle-shaped
N-methylphenazinium-tetra\-cyano\-quino\-dime\-tha\-ne (NMP-TCNQ).
The EPR spectrum of the powder (DPPH grains $< 1$~$\mu$m) shows
very narrow singlet lines with peak-to-peak linewidth,
\mbox{$W_{pp}\sim$0.15}~mT, having an almost constant $g$-value in
a wide temperature range from room temperature down to that of
liquid helium. On the other side, DPPH powder spread on the
restricted surface of the SMM crystal shows a broad EPR spectrum
around the resonance field, $B_0$, with characteristic shoulders
in the low temperature region (4--100~K) \cite{Rakvin2004}. Such a
spectrum represents a proportional ``fingerprint'' of the expected
field distributions at the area of the SMM crystal surfaces
covered by the probe. The relaxation of SMM magnetization in the
low temperature region (4--50~K) is significantly slower than the
detection time of EPR spectroscopy ($\tau >> t_m$)
\cite{Sessoli1993,Novak2005}. Thus, the effect of magnetic-field
fluctuation will be detected as the distribution of nearly static
magnetic field components. It was shown earlier
\cite{Rakvin2004,Rakvin2003} that the characteristic shoulders of
the complex powder spectrum of the probe depend on the SMM
orientation in the magnetic field. These shoulders exhibited
temperature-dependent shifts, which correlated with SQUID measured
magnetization of the SMM for the givens orientation in the
magnetic field. However, there is a weakness in the detection of
the shoulders, particularly for the largest shift, since its
intensity significantly decreases, below the limit of spectrometer
signal-to-noise ratio at low temperatures. In order to improve the
sensitivity of the method, a DPPH crystal is considered as a
better probe for the detection of the local shift as well as the
monitored distribution of the local fields. It is expected that
due to the presence of the fast electron spin exchange interaction
inside the DPPH crystal, the broad distribution of the local
fields at the surface of the SMM will be averaged into a single
line. In addition, it is expected that the obtained singlet will
be broader than the singlet of DPPH in the absence of the SMM
crystal. In the present work, the possibility of using a crystal
of DPPH as a local probe on the surface of the most frequently
investigated SMM, Mn$_{12}$-acetate,
[Mn$_{12}$O$_{12}$(CH$_3$COO)$_{16}$(H$_2$O)$_4]\cdot$2CH$_3$COOH$\cdot$4H$_2$O
(hereafter abbreviated as Mn12), will be investigated and
discussed.

\section{Experimental}

\begin{figure}
\centerline{\includegraphics[width=5cm,clip=]{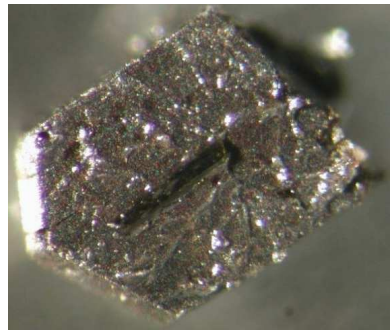}}
\caption{(Color online.) DPPH single crystal mounted by vacuum
grease on the surface of an Mn12 single crystal, shown under
$\sim$~20$\times$ magnification. The $a$~axis of the DPPH was
parallel to the $z$ axis of the Mn12. } \label{photography}
\end{figure}

Single crystals of Mn12 were grown using Lis' procedure
\cite{Lis1980}, and grew into rectangular rods of a few mm$^3$ in
dimension. The easy axis of magnetization, the $z$ axis, was along
the longest dimension of the crystals. Single crystals of DPPH
were crystallized from a solution of DPPH in ether using the
procedure described in \cite{Zilic2010}. The dimensions of the
DPPH crystals were approximately $1\times0.2\times0.2$~mm$^3$. The
crystals of DPPH were elongated along the crystallographic
$a$~axis. The DPPH crystal was fixed by using vacuum grease on the
Mn12 crystal, as shown in Fig.~\ref{photography}.

EPR measurements were carried out with a Bruker 580 FT/CW X-band
EPR spectrometer equipped with a standard Oxford Instruments Model
DTC2 Temperature Controller (\mbox{4--300~K}). All the
measurements were performed with a magnetic field modulation
amplitude of $\sim$~5~$\mu$T at 100~kHz.

\section{Results and discussion}

\begin{figure}
\centerline{\includegraphics[width=9cm,clip=]{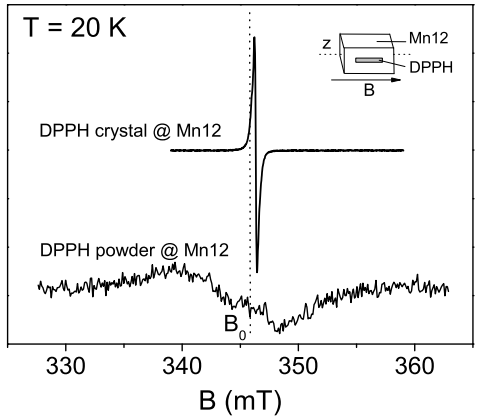}}
\caption{The EPR spectra of DPPH powder and a DPPH single crystal
deposited on the surface of an Mn12 crystal, at $T=20$~K. The
magnetic field was parallel to the easy axis of Mn12 ($B
\parallel z$). The dashed line represents the resonant field,
$B_0$, for both DPPH samples (without Mn12).}
\label{PowderCrystal}
\end{figure}

The EPR spectra of the single crystal of the DPPH probe at various
orientations in the magnetic field exhibit a narrow Lorentzian
profile of a detected singlet with an average $g$-value of
$g=2.004\pm0.001$ and peak-to-peak linewidths of
$W_{pp}=(0.15\pm0.03)$~mT, in the wide temperature range 4--300~K
\cite{Zilic2010}. The DPPH crystal was deposited on the Mn12
crystal with selected orientation, as shown in
Fig.~\ref{photography}. The $a$ axis of the DPPH crystal was
oriented parallel to the $z$ axis of the Mn12 (along the easy axis
of the Mn12). At room temperature, the probe mounted on the Mn12
crystal shows approximately the same spectra as the probe itself
at the resonance field, $B_0$. In the process of lowering the
temperature, the EPR line of the probe deposited on the Mn12
broadened and shifted from the resonance position $B_0$.
Fig.~\ref{PowderCrystal} illustrates significant differences in
the spectra between the powder and crystal DPPH deposited on the
Mn12 crystal and detected in the low temperature region (at 20~K).
One should note that the broad powder spectrum appeared from both
sides of the $B_0$ in contrast to the sharp crystal spectrum,
which exhibited a clear shift from the $B_0$.

\begin{figure}
\centerline{\includegraphics[width=9cm,clip=]{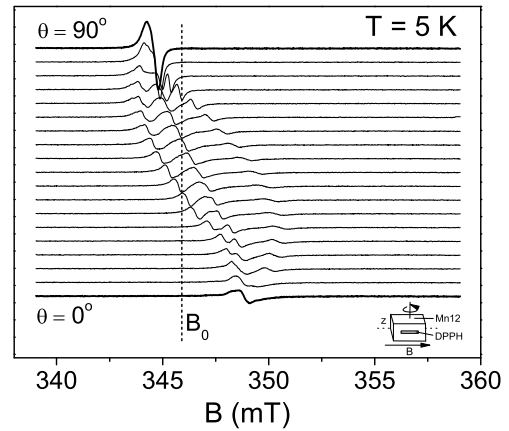}}
\caption{The EPR spectra of the DPPH single crystal deposited on
the surface of the Mn12 crystal, at $T=5$~K. The spectra were
recorded for a different angle, $\theta$, between magnetic field
$B$ and the $z$ axis of the Mn12 crystal, in steps of 5$^{\circ}$.
The dashed line represents the resonant field, $B_0$, for the DPPH
single crystal (without Mn12).} \label{DPPH1spectra}
\end{figure}

The EPR spectra for a different angle $\theta$ between magnetic
field $B$ and easy axis $z$, are shown in Fig.~\ref{DPPH1spectra}.

In the case when \mbox{$\theta =0^{\circ}$}, the detected spectral
line was shifted in the direction to a higher magnetic field, in
comparison to the $B_0$. For the perpendicular orientation of the
magnetic field (\mbox{$\theta =90^{\circ}$}) the line is shifted
in the opposite direction of the $B_0$. Between these two
canonical orientations, the spectral line exhibited an additional
splitting structure, as shown in Fig.~\ref{DPPH1spectra}. The
results obtained significantly differ from a similar experiment in
which a DPPH crystal was replaced by DPPH powder.

\begin{figure}
\centerline{\includegraphics[width=9cm,clip=]{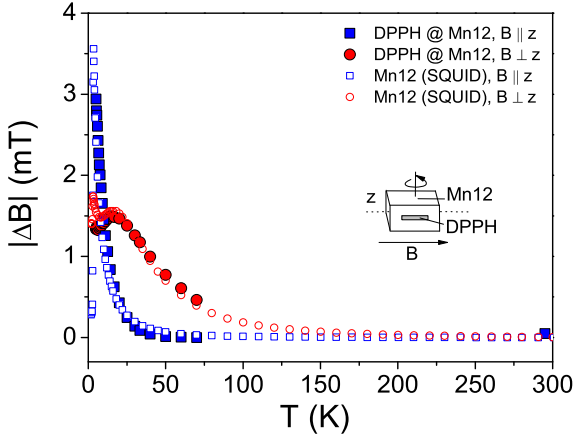}}
\caption{(Color online.) Comparison of EPR and SQUID magnetization
data. Temperature variation of the effective magnetic field
``experienced''by probes (DPPH crystal) on the surface of Mn12
measured by EPR (filled symbols) compared with the scaled SQUID
magnetization (open symbols).} \label{BvsT_DPPH1}
\end{figure}

As was demonstrated in Fig.~\ref{PowderCrystal} and in a series of
earlier experiments \cite{Rakvin2004}, a powder probe exhibits a
broad spectrum, with a pattern that contains characteristic
shoulders. As also shown earlier \cite{Rakvin2004}, these
shoulders are scaled with SQUID measured magnetizations (for two
orientations: \mbox{$B
\parallel z$} and \mbox{$B \perp z$}) in the same temperature interval. The same
procedure was applied on measured shifts for each orientation of
the Mn12 crystal labeled with a crystal probe. Indeed, the field
shift of the crystal probe for the corresponding orientation of
the Mn12 crystal shows proportionality with independently measured
SQUID magnetizations \cite{Rakvin2004}, as shown in
Fig.~\ref{BvsT_DPPH1}.

From a good match of EPR and SQUID data in Fig.~\ref{BvsT_DPPH1},
it is obvious that additional magnetic field ``experienced''by the
probes has the origin in the magnetization of Mn12. Larger shift
for \mbox{$B \parallel z$} compared to \mbox{$B \perp z$} is due
to much larger component of magnetization along the easy axis of
Mn12 compared to the hard axis. From the Fig.~\ref{BvsT_DPPH1} it
could be seen that magnitude of the averaged field shift for the
probe at the surface of Mn12 is $\sim$1~mT at low temperature.

In order to reveal the microscopic origin in the difference
between the crystal and  DPPH powder deposited on the Mn12
crystal, it was noted that different orientations of DPPH crystal
on the surface, for the same Mn12 crystal orientation in the
magnetic field, showed nearly the same shift of the spectrum line.
Thus, the results obtained support the above-mentioned expectation
that a wide distribution of local fields at the surface of the
Mn12 crystal is averaged out by the crystal probe (DPPH crystal).
At the same time, the ``long range'' of the exchange interaction
with an averaging behavior is absent for the powder probe. A
further consequence of this experiment is that the ``shorter
range'' of the spin-spin type of exchange interaction in powder
grains than in the crystal continuum emerges as the only possible
difference between the powder and crystal probe. It is expected
that a small particle within the powder could only partly average
local fields and finely a broad spectrum with characteristic
powder pattern is obtained. Additional important evidence should
be noted from the present experiment. Since, the exchange
interaction is expected to predominate on the level of the DPPH
molecule pairs \cite{Zilic2010,Fujito1981}, the distribution of
the local fields is expected to exhibit space inhomogeneity (with
possible variation of the gradient) at comparable distances. Such
local fields with space inhomogeneity at $\sim$1~nm scale are
indeed expected for Mn12 in the superparamagnetic state.

The nature of DPPH spectra (in the powder or crystal form) is
related to two dominant spin-spin interactions: dipole-dipole and
exchange interaction. A spin dipole-dipole interaction is
proportional to $1/a^6$ (where $a$ is the distance between
molecules) and contributes significantly to the broadening of the
resonance line since an unpaired electron spin density is present
at each DPPH molecule in the crystal. It is also known that such
broadening of the resonance line (assuming a Gaussian line shape)
could be approximated with the square root of the second moment of
the resonant line, $M_2$ \cite{Goldman1970}. In order to evaluate
the $M_2$ of DPPH ($S=1/2$), approximation calculus for a powder
sample with cubic symmetry was employed \cite{Poole1983}. The
second moment is additionally averaged over a sphere for spin $i$
to yield:
\begin{equation} \label{eM2}
M_2 =  \frac{10}{3}\frac{3}{5}g^{2}\beta^{2}S(S+1)\sum_j
r^{-6}_{ij}.
\end{equation}
Here, the term $i=j$ is excluded from the summation and the
symbols have their usual meanings. The factor $10/3$ (the
so-called ``$10/3$'' effect) is added because of subsidiary lines
that appear at $\sim 0, 2 g \beta H$ and $3 g \beta H$
\cite{Poole1983}. Considering the contribution of the nearest
molecules for the sphere of a radius of 15~\AA, the corresponding
dipolar contribution to the DPPH linewidth, $\Gamma_{d0}$, has
been calculated:
\begin{equation} \label{eDipolar}
\Gamma_{d0}\approx \sqrt{M_2}.
\end{equation}
where $\Gamma_{d0}$ is the half-width at half-intensity and is
given in Table~\ref{tParameters}.

\begin{table}
\begin{center}
\caption{Experimental and calculated parameters of linewidths for
the DPPH crystal with and without the Mn12 crystal, at 10~K. All
the quantities are expressed in mT.} \label{tParameters}
\resizebox{6cm}{!} {
\begin{tabular}{ccccc}

\hline
  $W_{pp0}$ & $W_{pp}$ & $\Gamma_{d0}$  & $H_{exch}$ & $\Gamma_{d}$ \\
 \hline
0.14 & 0.21 & 19.5 & 3142.8 & 23.9 \\
\hline
\end{tabular}}
\end{center}
\end{table}
Thus, the spin dipole-dipole contribution leads to a relatively
broad spectrum, which is two orders of magnitude broader than the
experimental linewidth value of DPPH ($\Gamma_{d0} >> W_{pp0}$).
The linewidths for the DPPH crystal with and without Mn12,
$W_{pp}$ and $W_{pp0}$, respectively, are given in
Table~\ref{tParameters}. The experimental spectral lines of the
probes showed approximately Lorentzian shapes, and the following
could be written for the effective Lorentzian linewidths
\cite{Weil1994}:
\begin{equation} \label{eHL}
\Gamma_{L0} = \sqrt{3} / 2 W_{pp0}.
\end{equation}

\begin{figure}
\centerline{\includegraphics[width=9cm,clip=]{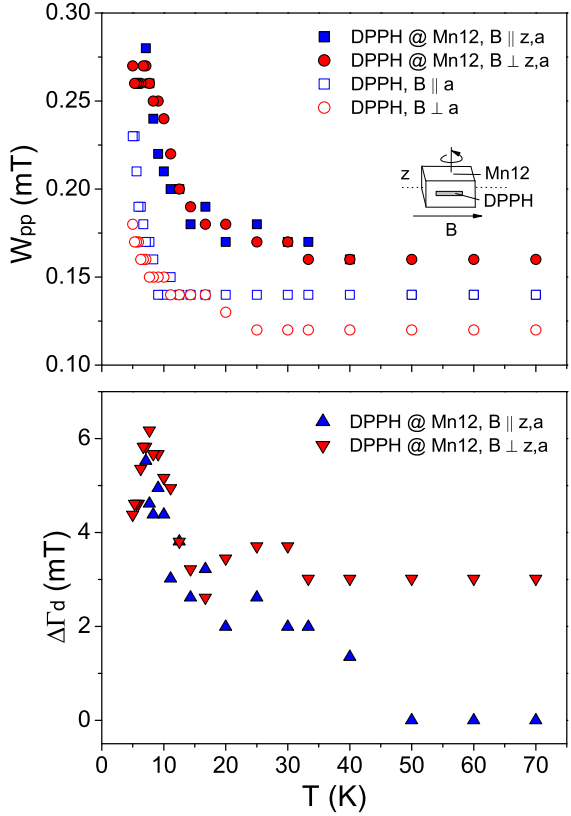}}
\caption{(Color online.) \emph{Upper:} Experimentally obtained
temperature variation of linewidths for DPPH (open symbols) and
for DPPH deposited on Mn12 (filled symbols). \emph{Lower:} The
calculated temperature dependence of the dipolar field, $\Delta
\Gamma_{d}$, caused by the Mn12 crystal at the position of the
DPPH crystal, according to relation \ref{eDeltaGamma}.}
\label{WvsT}
\end{figure}

In the case when the exchange interaction is strong, the
orientation of neighboring spins is exchanged at the rate of order
of $H_{exch}$ through mutual spin flips so that the local dipolar
field fluctuates at similar rate and tends to average out. In the
case if \mbox{$H_{exch}
>> \Gamma_{d0}$}, the broad line shape becomes nearly Lorentzian
with an approximate half-width \cite{Poole1983}:
\begin{equation} \label{eHexch}
\Gamma_{L0} \approx (\Gamma_{d0})^{2} / H_{exch}.
\end{equation}
Relation \ref{eHexch} and the data in Table \ref{tParameters} can
be used to estimate $H_{exch}$ for the DPPH crystal. The value
obtained is given in Table \ref{tParameters}
(\mbox{$H_{exch}=3142.8$~mT $\sim$ 4.2~K}) and is in good
agreement with earlier deduced values for the exchange constant of
DPPH \cite{Zilic2010}. If the DPPH crystal is deposited on the
Mn12 crystal, the local dipolar field of the DPPH molecule
increases by the contribution of the local Mn12 fields and the
same exchange interaction will reduce both dipolar contributions,
$\Gamma_{d}$, in an effectively Lorentzian line shape with a
peak-to-peak linewidth, $W_{pp}$. For experimentally measured
$W_{pp}$ one simply calculates the $\Gamma_{d}$ value, given in
Table~\ref{tParameters} (as in relation \ref{eHexch}). It is easy
to obtain ($\Delta \Gamma_d = \Gamma_{d}- \Gamma_{d0}= 4.4$~mT),
the average dipolar field being caused by the presence of the Mn12
crystal. This value also represents an average distribution of the
local fields on the surface of the Mn12 crystal at the position of
the DPPH crystal. The same procedure was applied for the
estimation of the linewidth of the Gaussian distribution of the
local fields measured on the surface of the Mn12 crystal at the
position of the DPPH crystal in a wide temperature interval
(Fig.~\ref{WvsT}):
\begin{equation}
\label{eDeltaGamma} \Delta \Gamma_d = \Gamma_{d0} (\sqrt{W_{pp} /
W_{pp0}}-1).
\end{equation}
%

%
%

%
Generally, these values indicate increases with decreasing
temperature in the monitored region of low temperatures. At this
moment, the corresponding activation process with the approximate
activation energy $\sim10$~K
 could not be simply correlated with
one of microscopic parameters known for the Mn12 crystal
\cite{Sessoli1993,Novak2005}. It is also important to note that
the obtained value of the Gaussian distribution of the local
fields coincides with the linewidth of the broad spectrum of DPPH
powder within an order of magnitude. This provides additional
indication that the ``long range'' spin-spin exchange interaction
present in the crystal is involved in the narrowing of the local
field distribution.

\section{Conclusions}

A DPPH crystal as a probe shows several advantages in respect to a
DPPH powder probe for detecting the distribution of local fields
on the surface of an SMM crystal. It is clearly shown that
``discontinuity'' (or ``short range'') of exchange interaction
significantly contributes to broaden the spectra of a powder
probe. This broadening is related to decreasing spectral intensity
and lead to uncertainty in the detection of the characteristic
spectral shoulders. Thus, temperature-dependent measurements,
especially at low temperature at higher local fields, are not
possible. On the other side, a crystal probe, due to the presence
of exchange interaction on the longer range scale than in powder,
efficiently averages all the local fields within the crystal,
resulting in a narrow Lorentz-type line around the center of
distribution, as seen in Fig.~\ref{PowderCrystal}. The narrowing
helps to estimate the shift of the spectral line more accurately,
even in the low temperature region. With this method, accuracy is
particularly improved for the detection of the average
distribution of the local fields in the form of the linewidth of
the Gaussian distribution. As shown above, this accuracy is
closely related to the accuracy of the exchange constant of the
probe. It should also be mentioned that the behavior of the
exchange interactions within organic-type paramagnetic crystals is
not known in detail. It is expected that these exchange
interactions depend on various microscopic models of exchange
between the paramagnetic molecules in the crystal, leading to
possible exchange distribution within the crystal continuum.
Indications for such symmetry-dependent distribution of the
exchange interaction within the crystal are present in the
splitting spectra of the DPPH probe when the magnetic field was
neither parallel nor perpendicular to the easy axis of Mn12
(Fig.~\ref{DPPH1spectra}). This characteristic of the exchange
interaction has also been noted for other organic crystal probes,
the DPPH crystal--containing solvent molecule (CS$_2$)
\cite{Zilic2010}, the
$\alpha$,$\gamma$-bisdiphenylene-$\beta$-phenylallyl (BDPA)
crystal and the previously studied TCNQ poly-crystal
\cite{Rakvin2005,Rakvin2003}. Thus, SMM crystals with a broad
distribution of local fields at the surface that exhibit space
inhomogeneity at the nano scale could be used as possible devices
for the investigation of the exchange interaction in organic-type
paramagnetic crystals. Indeed, such an application of SMM is under
investigation and the results obtained will be discussed and
published elsewhere.

\bigskip
\textbf{Acknowledgments}
\bigskip

This research was supported by the Ministry of Science, Education
and Sports of the Republic of Croatia (project 098-0982915-2939).
We are grateful to K.\,Mol\v{c}anov for help with depositing the
probe crystals on the Mn12 crystals; D.\,Paji\'{c} for SQUID
magnetization measurements and M.\,Juri\'{c} for the
crystallization of the DPPH crystals.

\bibliographystyle{h-physrev3}

\end{document}